\documentclass[aps,amsfonts,amsmath,prd,preprint,nofootinbib]{revtex4}
\usepackage{graphicx}

\newcommand{\beq}{\begin{equation}}
\newcommand{\eeq}{\end{equation}}

\def\lap{\lower.5ex\hbox{$\; \buildrel < \over \sim \;$}}
\def\gap{\lower.5ex\hbox{$\; \buildrel > \over \sim \;$}}

\begin{document}

\title{A quantum measure of the multiverse}

\author{Alexander Vilenkin}

\address{
Institute of Cosmology, Department of Physics and Astronomy,\\ 
Tufts University, Medford, MA 02155, USA}

\begin{abstract}

It has been recently suggested that probabilities of different events
in the multiverse are given by the frequencies at which these events
are encountered along the worldline of a geodesic observer (the
``watcher").  Here I discuss an extension of this probability measure to
quantum theory.  The proposed extension is gauge-invariant, as is the
classical version of this measure.  Observations of the watcher are
described by a reduced density matrix, and the frequencies of events
can be found using the decoherent histories formalism of Quantum
Mechanics (adapted to open systems).  The quantum watcher measure
makes predictions in agreement with the standard Born rule of QM.  

\end{abstract}

\maketitle

%

\section{Introduction}

A long-standing problem of inflationary cosmology is the so-called
measure problem.  In an eternally inflating universe, any event having
a non-vanishing probability will occur an infinite number of times,
and in order to assign probabilities to different events these
infinities must be regulated.  The problem is that the results turn
out to be highly sensitive to the choice of the regulator.  For
example, if one counts only events prior to some global time $t$, the
resulting probabilities are sensitive to the choice of the time
variable. (For a review of the measure problem, see, e.g.,
\cite{Freivogel}.)  An additional puzzle arises in the context of
quantum measurements.  Even if the regulator is specified, the
probabilities of different measurement results cannot be expressed as
expectation values of projection operators, as required by the Born
rule \cite{Page}.  The reason is that when the universe is so large
that it contains multiple copies of the experiment, one needs an
additional rule that would select a specific observer among the
identical copies (or assign probabilities to different copies \cite{Srednicki}).  Moreover, any acceptable probability measure should
satisfy the `correspondence principle', that is, its predictions
should agree with those of regular Quantum Mechanics in the range of
applicability of the latter.  Some of the currently popular measures
may satisfy this principle, at least under certain conditions, but
this has not been explicitly demonstrated. 

It has been recently suggested \cite{watcher} that the classical
measure problem can be naturally resolved by using a measure based on
a probe geodesic traversing the multiverse and encountering each type
of event an infinite number of times.  This geodesic can be thought of
as the worldline of an eternal observer, the "watcher". 
(It should be remembered though that the watcher is
not a physical entity and has no degrees of freedom of its own.)
Probabilities of different events are then identified with the
frequencies at which these events are encountered by the
watcher.\footnote{This measure prescription has close similarity to
  Nomura's single observer measure \cite{Nomura} and to the so-called
  "fat geodesic" measure \cite{fat}, but there are also important
  differences.  For a detailed discussion see \cite{watcher}.}    It
was shown in \cite{watcher} that the same probability distribution is
obtained for all geodesics, except for a set of measure zero.  A
crucial assumption underlying this prescription is that the geodesics
do not terminate.  In particular, it is assumed that big crunches in
the interiors of negative-energy (Anti-de Sitter) bubbles are
non-singular and are followed by bounces with subsequent expansion, so
that geodesics can be continued through the crunches.\footnote{If the
  vacuum landscape includes stable Minkowski vacua, then a generic
  geodesic eventually enters such a vacuum and stays there forever.
  Watcher's worldlines should then be selected from a measure-zero
  class of geodesics which do not get stuck in stable Minkowski
  vacua \cite{watcher}.} 

The purpose of this paper is to extend the watcher measure to include
quantum measurements.  I will argue that quantum probabilities in this
measure can be calculated using the formalism of decoherent histories
in Quantum Mechanics, adapted to open systems.  Furthermore, I will
show that the Born rule is consistent with the watcher measure and can
even be derived from it (modulo certain caveats). 

The paper is organized as follows.  In the next Section, I argue that
the horizon region of a geodesic observer in the multiverse should be
regarded as an open system interacting with its environment, in
contrast with a widely accepted view. In Section III, I review the
classical watcher measure and propose its quantum extension. A special
case where the effect of environment can be described by a Markov
stochastic process is considered in Section IV.  Consistency with the Born rule
is demonstrated in Section
V.  Finally, the conclusions of the paper are briefly summarized in
Section VI.

\section{Causal patch as an open system}


Observations of the watcher are confined to its causal patch, that is, to the spacetime region inside its future horizon.  In practice, a physical observer can monitor only a small fraction of the degrees of freedom in her causal patch.  The remaining degrees of freedom, both inside and outside the horizon, should be regarded as "the environment" and should be traced over.  The dynamics of the causal patch will  then be described by a reduced density matrix.  An important point, however, is that a density matrix description appears to be unavoidable, even if the observer monitors all the accessible degrees of freedom.  The reason is that there is a constant flow of information through the horizon to the exterior space.  As a result, the quantum state of the horizon region gets entangled with that of the exterior.  Since the part of spacetime outside the horizon cannot be observed, even in principle, the corresponding degrees of freedom should be traced over.  The horizon region should therefore be described by a reduced density matrix, even if the entire multiverse is in a pure quantum state \cite{BoussoSusskind}.

It has often been argued \cite{Banks1,Kleban,Nomura} that a causal
patch in a purely de Sitter landscape evolves like a finite closed system,
with the horizon acting as an impenetrable membrane, allowing no loss
of information.  It has also been suggested \cite{Banks} that a similar
picture should apply even in the presence of Anti-de Sitter vacua.
The causal patch would then reach a state of thermal equilibrium and
would remain in that state forever.  Apart from very rare large
thermal fluctuations, this state would not exhibit any arrow of
time.
I will now briefly summarize some of the arguments for and against this equilibrium picture of de Sitter space, but I will say at the outset that in the present paper I will not subscribe to this view and assume instead that the causal patch interacts with the environment and evolves as an open system.  

In classical GR, a causal patch in de Sitter space can be described using a static coordinate system,
\beq
ds^2 = (1-H^2 r^2)dt^2 - (1-H^2r^2)^{-1}dr^2 - r^2 d\Omega^2.
\eeq
Timelike and null geodesics approach the horizon $r=H^{-1}$ in the asymptotic future, but do not leave the causal patch in a finite time $t$.  Moreover, no timelike or null geodesic can enter the causal patch from outside.
This seems to suggest that the causal patch is indeed a closed system.  However, things may be different in quantum theory.  A detector at $r=0$ detects a thermal spectrum of particles emanating from the horizon.  These particles can be thought of as being produced in pairs, with the two members of the pair being on opposite sides of the horizon.  As a result the particles within the causal patch are entangled with the particles outside.

It has been recently suggested \cite{Mathur,Braunstein,Polchinski} that the semiclassical spacetime geometry may be drastically modified in the vicinity of black hole horizons, resulting in "fuzzballs" or "firewalls".  The firewalls absorb and re-emit all incoming information, so the evolution of the region outside the horizon is unitary.
One might expect that a similar picture could apply to de Sitter space, but the problem is that de Sitter horizons are observer-dependent.  Spacetime distortion and firewalls near the horizon appear to be inconsistent with the fact that all points in de Sitter space are equivalent.



As pointed out in \cite{watcher}, an interesting situation arises in multiverse models with a purely de Sitter landscape, where all vacua have positive energy density. The semiclassical transition rate from vacuum $j$ to vacuum $i$, detected by an inertial observer, is then given by \cite{CdL}
\beq
\kappa_{ij}=(4\pi/3)H_j^{-3}\Gamma_{ij} ,
\label{kappa}
\eeq
where $H_j=(8\pi\rho_j/3)^{1/2}$ is the de Sitter expansion rate in vacuum $j$ (in Planck units), $\rho_j$ is the corresponding vacuum energy density, and $\Gamma_{ij}$ is the nucleation rate per unit spacetime volume for bubbles of vacuum $i$ in parent vacuum $j$, 
\beq
\Gamma_{ij}=A_{ij} e^{-I_{ij}-S_j}.
\label{Gamma}
\eeq
Here, $I_{ij}$ is the Euclidean action of the tunneling instanton, $A_{ij}$ is a prefactor arising from integration of small perturbations around the instanton, and $S_j$ is the entropy of vacuum $j$. 
The instanton action and the prefactor $A_{ij}$ are symmetric with respect to interchange of $i$ and $j$ \cite{LeeWeinberg}.  Hence, we can write
\beq
\kappa_{ij}/\kappa_{ji} = (H_i/H_j)^3 \exp(S_i -S_j).
\label{balance}
\eeq

The quantity $e^{S_j}$ has the interpretation of the number of (accessible) microstates in a horizon region of vacuum $j$, and the relation
\beq
\kappa_{ij}/\kappa_{ji} \propto \exp(S_i -S_j)
\label{balance2}
\eeq
can be thought of as expressing the detailed balance condition, which is necessary for equilibrium (microcanonical) distribution to establish \cite{Vanchurin,symmetree}.  However, the prefactor in (\ref{balance}) violates the detailed balance.  
Even if this violation is small, it indicates that the equilibrium picture of de Sitter space can only be approximate.\footnote{One might think that quantum gravity corrections to the entropy might compensate for the prefactor $H^{-3}$, thus restoring detailed balance. However, the factor $e^{S_j}$ in (\ref{Gamma}) is shorthand for
the semiclassical path integral around the Euclidean de Sitter saddle point corresponding to
the parent vacuum and already includes quantum corrections
\cite{watcher}.}  

Furthermore, in the presence of Anti-de Sitter vacua, it was argued in Ref.~\cite{watcher} that strong violations of detailed balance are likely to occur at bounces replacing the big crunch singularities in Anti-de Sitter bubbles.  Because of the high energy densities reached near the bounce, the crunch regions are likely to be excited above the energy barriers between different vacua,  so transitions to other vacua are likely to occur \cite{Piao,Jun,Param}.  The duration of the bounce, however, is very short, so there is no time for the region to explore its available phase space and reach thermal equilibrium.  In particular, there seems to be no reason for transition probabilities from the bounce region to different de Sitter vacua to be related to the entropies of those vacua.  Such violations of ergodicity may be responsible for the observed arrow of time.   

Realizing that these issues are far from being settled, here we shall adopt the following assumptions and explore heir consequences for the measure problem.
(1) Anti-de Sitter bounces do occur and are accompanied by strong
violations of ergodicity.  The bounces allow a semiclassical
description, so the watcher's geodesic can be continued through the
bounces.\footnote{A "geodesic" is a classical concept that can only be
  defined in a semiclassical spacetime background.  Some approaches to
  quantum gravity, in particular the holographic ideas, suggest that a
  full quantum description should be in terms of the wave function (or
  density matrix) of a region encompassed by an apparent horizon
  surface (e.g., \cite{Nomura,BoussoSusskind,census}). Then geodesics
  representing possible trajectories of a watcher may play no
  fundamental role, except perhaps in some appropriate limit.} 
   (2) The causal patch of the watcher interacts with its environment; its evolution can be described by a reduced density matrix. 

\section{The watcher measure and its quantum extension}

\subsection{The classical watcher measure}

In the classical version of the watcher measure \cite{watcher}, an event $A$ is counted if the watcher's geodesic crosses the spacetime domain $D_A$ of the event.  The domain $D_A$ is defined as the minimal spacetime region necessary to distinguish this type of event from others.  In order to account for the different sizes of the domains, the number of encountered events is then renormalized by a factor $\sigma_A^{-1}$, where $\sigma_A$ is the cross-section (having the dimension of 3-volume) that the domain $D_A$ presents to the watcher's geodesic.  

Here I am going to use a slightly different, but essentially equivalent prescription.  For each domain $D_A$ we shall define a point that we shall call its center.  If $D_A$ is small enough, so that spacetime curvature in $D_A$ can be neglected, then we can define its center by analogy with the center of mass: in the standard Minkowski coordinates, the center is a point $x_A^\mu$, such that 
\beq
\int_{D_A} d^4 x (x^\mu - x_A^\mu) = 0.
\eeq
This definition can be generalized to curved spacetime as
\beq
\int_{D_A} d^4 x \sqrt{-g} v^\mu(x,x_A) = 0 ,
\eeq
where $v^\mu(x,x_A)$ is a vector at point $x_A$ pointing in the direction of the (shortest) geodesic connecting the points $x_A$ and $x$ and having magnitude proportional to the length of that geodesic.  In fact, the precise definition of the center is unimportant, as long as it is a well specified point in the domain $D_A$.\footnote{The choice of a center may be important in cases where the event $A$ represents a ``story" whose domain extends over more than a Hubble time in the time direction.  This leads to the Guth-Vanchurin paradox \cite{GuthVanchurin}, and the probabilities will depend on whether we choose the center at the beginning or at the end of the story.}

We shall adopt the prescription that an event of type $A$ is counted whenever the center of its domain lies within a specified small distance range $\epsilon$ from the watcher's geodesic.  If this condition is satisfied, we shall say that the event has been encountered by the watcher.  The precise value of $\epsilon$ is also unimportant, as long as it is sufficiently small.  This prescription is essentially the same as that in the fat geodesic measure \cite{fat}, except that we assume that $\epsilon$ is smaller than the domain $D_A$, so no more than one event can be encountered at a time.  

Let $t=0$ be an arbitrary point on the watcher's worldline and $N_A(T)$ the number of events of type $A$ encountered during the time interval $0<t<T$.  The relative probability of events $A$ and $B$ is then identified with the relative frequency of these events as they are encountered by the watcher,
\beq
\frac{p_A}{p_B} = \lim_{T\to\infty} \frac{N_A (T)}{N_B (T)} .
\label{pApB}
\eeq
In this formulation, no renormalization of the numbers of events is required.  It is also clear that the resulting probabilities are independent of the choice of the time variable $t$, as long as it is monotonic along the watcher's geodesic.

\subsection{Decoherent histories for an open system}

An extension of the watcher measure to quantum theory is most naturally obtained using the decoherent histories formulation of Quantum Mechanics \cite{Griffiths,Omnes,HGM}.  Possible histories of the watcher can be represented by chains of projection operators at a sequence of times, $0< t_1 < t_2 < ... < T$~,
\beq
h:~ P_{i_1},~ ... ~P_{i_n} ,
\label{h}
\eeq
where $t_n=T$ and the subscript $i_k$ indicates the alternative that has been chosen at time $t_k$ in the particular history $h$.\footnote{More generally, one can consider a chain of projectors $P^{(k)}_{i_k}$,
where the label $(k)$ allows for the possibility of choosing different sets of projectors at different times.  Our discussion can be straightforwardly extended to this case.}  The projectors $P_{i_k}$ act in the Hilbert space of the watcher's causal patch; the watcher itself is completely specified by its geodesic and has no independent degrees of freedom of its own. 
 
We assume that the projectors $P_{i_k}$ belong to an exhaustive and mutually exclusive set,
\beq
\sum_i P_i = 1,
\eeq
\beq
P_i P_j = P_i \delta_{ij}.
\eeq
We shall think of these projectors as representing records of events that happened in the time interval $t_{k-1}<t<t_k$, rather than the events that occurred at the moment $t_k$.  For example, if some measurement was made in this time interval, then $P_{i_k}$ are the projectors on the possible outcomes of the measurement. 

For any two histories $h$ and $h'$, we can define the decoherence functional \cite{HGM}
\beq
D(h',h)=Tr \left( P_{i_n} K_{t_{n-1}}^{t_n} \left[ P_{i_{n-1}} ... K_{t_1}^{t_2} \left[ P_{i_1} K_0^{t_1} [\rho(0)] P_{{i'}_1} \right] ... P_{{i'}_{n-1}} \right] P_{{i'}_n} \right)
\label{Dhh'}
\eeq
Here, the `super-operators' $K_{t_i}^{t_{i+1}}$ evolve the density matrix from $t_i$ to $t_{i+1}$,
\beq
K_{t_i}^{t_{i+1}}[\rho(t_i)] \equiv e^{-i{\cal H}(t_{i+1} -t_i)} \rho(t_i) e^{i{\cal H}(t_{i+1} - t_i)} =\rho(t_{i+1}) ,
\label{K}
\eeq
$\rho(0)$ is the initial density matrix at $t=0$ and ${\cal H}$ is the Hamiltonian.

We say that histories $h$ and $h'$ decohere if $D(h',h) \approx 0$.  If all histories in the set (\ref{h}) decohere, then each history can be assigned a probability,
\beq
p(h) = D(h,h).
\label{ph}
\eeq
Histories generally decohere when the time intervals $\Delta t_k = t_k-t_{k-1}$ are increased and when the history is coarse grained (that is, when the projectors are bunched together into a smaller set of projectors).
With a suitable choice of basis (the so-called `pointer basis'), decoherence can be achieved even with minimal coarse-graining and for a very small time separation between successive events \cite{Zurek}.  

In the case of an open system, the universe is divided into the
`system of interest' ${\cal S}$ and the `environment' ${\cal E}$.  The
histories of the system are then specified by projectors of the form
$P=I_{\cal E} \otimes P_{\cal S}$, where $I_{\cal E}$ is the identity
in the Hilbert space of the environment and $P_{\cal S}$ is a
projector in the Hilbert space of the system.  However, the trace in
Eq.~(\ref{Dhh'}) for the decoherence functional should still be taken
over the full Hilbert space, including both the system and the
environment, and the Hamiltonian ${\cal H}$ in the evolution operators in (\ref{K}) is the full Hamiltonian.  It is not generally possible to express the decoherence functional only in terms of the reduced density matrix of the system,
\beq
{\tilde\rho}(t) = Tr_{\cal E}\rho(t).
\label{reduced}
\eeq
The reason is that the evolution of the system is generally influenced by the correlations that develop between the system and the environment.  We expect, however, that such correlations should be unimportant for the causal patch of the watcher.  For example, the members of particle-antiparticle pairs outside the de Sitter horizon are quickly driven away by the de Sitter expansion, and it seems reasonable to assume that they have little effect on the subsequent evolution of the horizon interior.\footnote{In models of inflation, regions outside the apparent horizon can later become observable.  In such cases super-horizon correlations can be significant, but it appears that correlations beyond the true causal horizon can still be neglected.}  Assuming this to be the case, it should be possible to define the evolution operator ${\tilde K}_{t_i}^{t_{i+1}}$ for the reduced density matrix \cite{Zurek}   
\beq
{\tilde K}_{t_i}^{t_{i+1}}[{\tilde\rho}(t_i)]  = {\tilde\rho}(t_{i+1}) .
\eeq
The decoherence functional is then given by
\beq
D(h',h)=Tr_{\cal S} \left( P_{i_n} {\tilde K}_{t_{n-1}}^{t_n} \left[ P_{i_{n-1}} ... {\tilde K}_{t_1}^{t_2} \left[ P_{i_1} {\tilde K}_0^{t_1} [{\tilde\rho}(0)] P_{{i'}_1} \right] ... P_{{i'}_{n-1}} \right] P_{{i'}_n} \right) .
\label{tildeDhh'}
\eeq
It should be noted that the evolution operator ${\tilde K}_{t_i}^{t_{i+1}}$ for an open system is not generally given by Eq.~(\ref{K}).

\subsection{The quantum watcher measure}

In order to extend the watcher measure to quantum theory, we shall assume that the set of projection operators 
$P_{i_k}$ includes projectors on the states corresponding to all types of events of interest.
The number of events of type $A$ in a history $h$ defined by the chain of projectors (\ref{h}) is then 
\beq
N_A (h;T)=\delta_{i_1,A}+\delta_{i_2,A}+... ~,
\label{NA}
\eeq
and the average number of such events in the time interval $0<t<T$ is given by the sum over histories
\beq
N_A (T) = \sum_h p(h) N_A (h;T).
\label{NA(T)}
\eeq
As in the classical version of the measure, we shall identify the relative probability of events $A$ and $B$ with their relative frequency,
\beq
\frac{p_A}{p_B} = \lim_{T\to\infty} \frac{N_A (T)}{N_B (T)} ,
\label{pApB2}
\eeq
where $N_A$ and $N_B$ are now given by (\ref{NA(T)}).  We assume that the time intervals $\Delta t_i = t_i - t_{i-1}$ are sufficiently small, so that no relevant events are missed between the sampling times $t_i$.  For example, one can choose $\Delta t_i$ to be somewhat larger than the characteristic time of decoherence.  We also assume that the typical time separation $\Delta t_i$ is kept fixed in the limit $T\to\infty$, so the number of sampling moments $n$ becomes infinite in the limit.

Assuming that the vacuum landscape is irreducible, that is, that any vacuum can be reached from any other vacuum in a finite number of transitions, the initial state at $t=0$ will eventually be completely forgotten, and the asymptotic frequencies of events will be determined entirely by the properties of the landscape.  The relative probability (\ref{pApB2}) should therefore be independent of the initial density matrix ${\tilde\rho}(0)$ (which appears in the definition of $p(h)$; see Eqs.~(\ref{ph}),(\ref{tildeDhh'})).  One can, for example use some pure state $|\psi\rangle\langle\psi|$ or the asymptotic density matrix ${\tilde\rho}(\infty)$.

\subsection{Timestep evolution}

Suppose we want to find relative probabilities for some set of events, labeled by index $J=1,2,...~$.  To distinguish these events of interest from other, irrelevant events, we shall refer to them as ``marked events".
We shall consider a set of alternative histories specified by different sequences of marked events, $J_1, J_2, ...~$, without specifying the times at which the events occurred.  These histories can be thought of as representing branching Everett's worlds, with the branching points corresponding to marked events.

Next, we define a branching ratio $T_{IJ}$ as the probability to observe event $I$ after observing event $J$ (without any marked events in between).  This can be found as 
\beq
T_{IJ}=\sum_J^I p(h) ,
\label{TIJ}
\eeq
where the summation is over all histories $h$ starting at $J$ and
encountering $I$ before any other marked event.\footnote{We are
  usually interested in macroscopic events, represented by a large
  number of microscopically indistinguishable states.  Such states are
  characterized by density matrices which may depend on prior events.
  Here I ignore this complication and assume that each marked event
  corresponds to a pure state.}  (Note that we do not assume that all
histories in the sum (\ref{TIJ}) encounter event $I$ at the same time.)
    From the definition of the branching ratios it is clear that they satisfy
\beq
\sum_I T_{IJ} = 1
\eeq
and that they do not depend on the choice of the time variable $t$.  

We now introduce a discrete timestep variable $n$, which takes integer values, $n=0, 1, 2, ...$, and which is incremented by one at every branching transition.  Given a probability distribution $p_J(n)$ at step $n$, the distribution at step $(n+1)$ can be found from the equation
\beq
p_I(n+1)= \sum_J T_{IJ} p_J(n).
\label{discevol1}
\eeq

The distribution $p_I(n)$ can be thought of as describing an ensemble of branching histories, with each history including $n$ events.  In the limit $n\to\infty$, $p_I(n)$ approaches a stationary distribution $p_I^{(\infty)}$ satisfying
\beq
p_I^{(\infty)}= \sum_J T_{IJ} p_J^{(\infty)}.
\label{stationary}
\eeq
This has a unique solution, assuming that the marked set of events is irreducible, that is, that it does not split into subsets which cannot be accessed from one another \cite{watcher}.  Since $T_{IJ}$ do not depend on the choice of $t$, the solution $p_I^{\infty}$ should also be gauge-independent.  The standard relation between the ensemble and time averaging implies that $p_I^{(\infty)}$ should be equal to the frequency at which event $I$ is encountered by the watcher.  We therefore expect the frequencies of events found from Eq.~(\ref{stationary}) to agree with those found from (\ref{pApB2}).


\section{Markovian evolution}

\subsection{The Lindblad equation}

As an illustration, we shall now consider a simplified model, where the reduced density matrix ${\tilde\rho}(t)$ can be represented as
\beq
{\tilde\rho}(t) = \sum_j p_j(t) |j\rangle \langle j| .
\label{rhot}
\eeq
Here, $|j\rangle$ are Schrodinger state vectors,
\beq
i\frac{\partial}{\partial t} |j\rangle = {\tilde{\cal H}}|j\rangle ,
\label{Schrodinger}
\eeq
which are assumed to form an orthonormal basis in the Hilbert space, and ${\tilde{\cal H}}$ is the Hamiltonian of the watcher's system (the causal patch).  The quantity $p_j(t)$ has the meaning of the probability to find the system in state $j$, and
\beq
\sum_j p_j(t) = 1 .
\eeq
For a closed system we would have $p_j={\rm const}$, but we shall assume that the Hamiltonian evolution (\ref{Schrodinger}) is punctuated by transitions caused by interaction with the environment, resulting in time variation of $p_j$.
We shall also assume that the evolution of $p_j(t)$ is Markovian\footnote{Markovian means that the evolution has no memory, so that ${\dot{\tilde\rho}}(t)$ depends on ${\tilde\rho}$ at time $t$, but not at earlier times.} and is described by the rate equation
\beq
{\dot p}_i = \sum_j (\kappa_{ij} p_j - \kappa_{ji} p_i)	,
\label{rateeq}
\eeq
where $\kappa_{ij}$ are the corresponding transition rates.  These assumptions should apply if the transitions are due to bubble nucleation in the multiverse; then Eq.~(\ref{rateeq}) is the standard rate equation \cite{GSVW} with $\kappa_{ij}$ given by (\ref{kappa}) (assuming that the time variable $t$ is the proper time along the watcher's geodesic).  More generally, the assumptions may give a reasonable approximation if the basis $|i\rangle$ is chosen to be the ``pointer basis", in which decoherence occurs on a very short timescale \cite{Zurek}.  I will later indicate how the assumptions can be relaxed.

Differentiating Eq.~(\ref{rhot}) with respect to $t$ and using Eqs.~(\ref{Schrodinger}),(\ref{rateeq}), we obtain
\beq
{\dot{\tilde\rho}}(t) = -i[{\tilde {\cal H}},{\tilde\rho}] + \sum_{i,j} \kappa_{ij} p_j(|i\rangle \langle i| - |j\rangle \langle j| ).
\label{rhoeq}
\eeq
Introducing the operators
\beq
Q_{ij}=|i\rangle \langle j| ,
\label{Q}
\eeq
\beq
Q_{ij}^\dagger=|j\rangle \langle i| = Q_{ji},
\label{Qdagger}
\eeq
we have
\beq
Q_{ij}^\dagger {\tilde\rho} (t) Q_{ij} = p_i(t)|j\rangle \langle j| ,
\eeq
\beq
Q_{ij} Q_{ij}^\dagger = |i\rangle \langle i| ,
\eeq
and
\beq
Q_{ij} Q_{ij}^\dagger {\tilde\rho}(t) = p_i (t) |i\rangle \langle i| = {\tilde\rho}(t) Q_{ij} Q_{ij}^\dagger .
\eeq
With the aid of these relations, we can rewrite Eq.~(\ref{rhoeq}) as
\beq
{\dot{\tilde\rho}} = -i[{\tilde{\cal H}},{\tilde\rho}] -\frac{1}{2}\sum_{i,j} \kappa_{ij} \left( Q_{ij}^\dagger Q_{ij}{\tilde\rho} +  Q_{ij}^\dagger{\tilde\rho} Q_{ij} - 2 Q_{ij}{\tilde\rho} Q_{ij}^\dagger \right) \equiv  -i[{\tilde{\cal H}},{\tilde\rho}] + {\cal L}{\tilde\rho}.
\label{Lindblad1}
\eeq

An equation of the form (\ref{Lindblad1}) is known as the Lindblad equation \cite{Lindblad}.
This is the general form of a linear, Markovian evolution equation for the density matrix that preserves its unit trace and positivity.  These properties hold for arbitrary operators $Q_{ij}$.
With $Q_{ij}$ of the form (\ref{Q}), the Lindblad equation
(\ref{Lindblad1}) has no more content than the rate equation
(\ref{rateeq}), but it may be useful in a more general context, when
the ansatz (\ref{rhot}) for the density matrix is not imposed.
Nonlocal generalizations of Eq.~(\ref{Lindblad1}), where the evolution
is non-Markovian, have also been discussed (e.g., \cite{nonlocal}).  A
method for solving the Lindblad equation (\ref{Lindblad1}) in an
operator form has been given in \cite{Nakazato}.

The microcanonical distribution,
\beq
{\tilde\rho} \propto  {I},
\label{microcanonical}
\eeq
where ${I}$ is a unit operator, is a solution of Eq.~(\ref{Lindblad1}), provided that
\beq
\sum_{i,j} \kappa_{ij}[Q_{ij},Q_{ij}^\dagger] = 0.
\label{ergocond}
\eeq
The latter condition is satisfied, for example, when $Q_{ij}^\dagger = Q_{ji}$ and $\kappa_{ij} = \kappa_{ji}$.  If (\ref{microcanonical}) is a solution, then this solution is approached asymptotically at late times.
According to our discussion in Sec.~II, we expect (\ref{ergocond}) to be weakly violated in a purely de Sitter landscape and to be strongly violated in the presence of Anti-de Sitter bounces.  In the general case, since $\kappa_{ij}$ and $Q_{ij}$ do not have explicit time dependence, we expect $\tilde\rho(t)$ to approach asymptotically a stationary distribution with 
\beq
[{\tilde {\cal H}},{\tilde\rho}(t\to\infty)]=0 , ~~~~~ {\cal L}{\tilde\rho}(t\to\infty) = 0.
\eeq

\subsection{Gauge-independence}

The quantities $\kappa_{ij}$ in the rate equation (\ref{rateeq}) are transition rates per unit time, and their magnitude depends on one's choice of the time variable.  If $\tau$ is the proper time along the watcher's geodesic, we can introduce a new variable $t$ as
\beq
dt = H^\beta d\tau,
\label{ttau}
\eeq
where $H$ is the Hubble expansion rate.  (For $\beta=1$, $t$ is the scale factor time.)  The transition rates in the new time variable are then related to the proper time rates by
\beq
\kappa_{ij}^{(\beta)}=H_j^{-\beta}\kappa_{ij}^{(0)},   
\label{kappabeta}
\eeq
so the rate equation can be rewritten as
\beq
{\dot p}_i = \sum_j (\kappa_{ij}^{(0)}H_j^{-\beta} p_j - \kappa_{ji}^{(0)}H_i^{-\beta} p_i)	,
\label{rateeqbeta}
\eeq
Solutions of this equation for $\beta\neq 0$ will clearly be different from the $\beta=0$ proper time solutions.  This applies in particular to the stationary distribution, which is approached in the limit $t\to\infty$,
\beq
p_j^{(\beta)}(t\to\infty)= H_j^\beta p_j^{(0)}(t\to\infty).
\eeq
\label{pbeta}
The density matrices for different values of $\beta$ will therefore also be different.

This gauge-dependence is not surprising.  The density matrix ${\tilde\rho}(t)$ describes an ensemble of watchers at a given value of the global time $t$, and the gauge-dependence has the same origin as in a global time cutoff.  An important point, however, is that the branching ratios (\ref{TIJ}) which appear in Eq.~(\ref{stationary}) for the probabilities are gauge-independent.  The gauge-independence of $T_{IJ}$ can be verified explicitly in the case where the marked set of `events' coincides with the complete set of states $|i\rangle$ in the pointer basis. 
Then we have
\beq
T_{ij}=\frac{\kappa_{ij}}{\sum_k \kappa_{kj}},
\label{T}
\eeq
and the gauge-dependent factor in Eq.~(\ref{kappabeta}) for $\kappa_{ij}^{(\beta)}$ cancels out.

\section{The Born rule}

\subsection{Consistency with Quantum Mechanics}

We shall now verify that our measure prescription agrees with the usual rules of quantum mechanics.   That is, that the probabilities of different measurement results for a quantum system are given by the standard Born rule.  The following argument is essentially the same as in Refs.~\cite{Finkelstein,Hartle,Mukhanov,Wada,Aguirre}, except that these papers consider an ensemble of identical experiments,\footnote{It has been noted in \cite{Hartle} that the argument can also be applied to repeated measurements on a single system.} while we are interested in experiments encountered at different times along the watcher's worldline.  

As a simple example, we shall consider an experiment measuring the spin projection of a spin-1/2 particle on a given axis.  Disregarding for a moment the environment degrees of freedom, the quantum state prior to the measurement is
\beq
|\psi\rangle_{\rm before} = \left( \alpha |\uparrow \rangle + \beta |\downarrow\rangle\right) |A_r\rangle ,
\label{before}
\eeq
where $|\uparrow\rangle$ and $|\downarrow\rangle$ are respectively the states with spin in the ``up" and ``down" directions along the axis, $\alpha$ and $\beta$ are complex coefficients satisfying $|\alpha|^2 +|\beta|^2 = 1$,  and $|A_r\rangle$ is the ``ready" quantum state of the measuring apparatus.  After the spin and the apparatus are allowed to interact, the spin state is entangled with that of the apparatus,
\beq
|\psi\rangle_{\rm after} = \alpha |\uparrow \rangle |A_\uparrow\rangle  + \beta |\downarrow\rangle |A_\downarrow\rangle ,
\label{after}
\eeq
where $|A_\uparrow\rangle$ and $|A_\downarrow\rangle$ are the states of the apparatus corresponding to ``spin up" and ``spin down" measurements, respectively.  Shortly afterwards, interaction with the environment causes the superposition in (\ref{after}) to decohere.  At that point the measurement is over, and its result is represented by the density matrix, obtained by tracing over all degrees of freedom except those of the apparatus, 
\beq
\rho_A = |\alpha|^2 |A_\uparrow\rangle \langle A_\uparrow | + |\beta|^2 |A_\downarrow\rangle \langle A_\downarrow | .
\label{rhoA}
\eeq

Like any other process, the above described measurement will be observed by the watcher an infinite number of times.  The successive observations will be separated by enormous time intervals.  If the watcher's world line passes near the center of a spin measurement experiment in some habitable bubble, it is rather unlikely to hit an identical experiment in the same bubble.  The world line will move on into the multiverse, crossing a number of different bubbles.  It will eventually return to the quantum state (\ref{before}) on a timescale which is set by bubble nucleation rates (and is typically very large).  Let us consider histories including, apart from other irrelevant events, a sequence of $N$ spin measurements at times $t_1, t_2, t_3, ..., t_N$.  There are $2^N$ distinct histories, represented by chains of projection operators of the form
\beq
h:~ P_{\uparrow} ... P_{\downarrow} ... P_{\downarrow} ...~,
\label{Cupdown}
\eeq
where $P_{\uparrow}$ and $P_{\downarrow}$ are projectors onto the states $|A_\uparrow\rangle$ and $|A_\downarrow\rangle$ of the measuring device.  The probability of a given history is
\beq
p(h) = \left( |\alpha|^2\right)^{N_\uparrow} \left(|\beta|^2\right)^{N_\downarrow} ,
\label{ph2}
\eeq
where $N_\uparrow$ and $N_\downarrow$ are respectively the numbers of up and down spin measurements in that history.  Here I assume that all measurements are uncorrelated, which is justified, considering that the system completely forgets its initial state on the recurrence timescale.  

The probabilities in Eq.~(\ref{ph2}) are normalized so that
\beq
\sum_h p(h) = 1,
\eeq
where the summation is over the histories (\ref{Cupdown}).  (Note that the measurement times $t_1, t_2, ...$ are the same for all histories.)  There is a factor $|\alpha|^2$ for each spin up and a factor $|\beta|^2$ for each spin down measurement in (\ref{ph2}).  It follows that the probability of having a given number $N_\uparrow$ 
of spin up measurements is given by a binomial distribution \cite{Mukhanov},
\beq
p_N(N_\uparrow) = \binom{N}{N_\uparrow} \left( |\alpha|^2\right)^{N_\uparrow} \left(|\beta|^2\right)^{N-N_\uparrow}.
\label{binomial}
\eeq
Let $f_\uparrow = N_\uparrow/N$ be the fraction of measurements that gave spin up.  Its mean value is then
\beq
\langle f_\uparrow\rangle = |\alpha|^2 ,
\eeq
and its variance is
\beq
\delta f_\uparrow \equiv \left( \langle f^2_\uparrow \rangle - \langle f_\uparrow\rangle^2 \right)^{1/2}  = \frac{|\alpha\beta|}{\sqrt{N}} .
\eeq
In the limit $N\to\infty$, we have $\delta f_\uparrow \to 0$, and the distribution approaches  a delta-function,
\beq
p_N(f_\uparrow) \to \delta (f_\uparrow - |\alpha|^2).
\label{delta}
\eeq

Applied to the watcher measure, this means that the frequencies of spin up and spin down measurements are precisely given by the Born rule for all watcher histories, except a set of measure zero.
It can be shown that the same conclusion applies to measurements that can have more than two different outcomes \cite{Finkelstein,Hartle,Wada,Aguirre} and to situations where the states of the macroscopic measuring device corresponding to different outcomes include large groups of macroscopically indistinguishable microstates \cite{Aguirre}.

\subsection{`Derivation' of the Born rule}

It has been suggested in Refs.~\cite{Finkelstein,Hartle,Mukhanov,Wada,Aguirre}
(see also \cite{Knobe}) that a slight modification of the above analysis
can be regarded as a {\it derivation} of the Born rule.  Here is a
rough sketch of the argument, adopted to our case. 

A sequence of spin measurements observed by the watcher can be represented by a density matrix
\beq
\rho = \rho^{(1)}_A \otimes \rho^{(2)}_A \otimes ... \otimes \rho_A^{(N)} ,
\label{rhoAN}
\eeq
with $\rho^{(k)}_A$ corresponding to the measurement at time $t_k$.  All of $\rho^{(k)}_A$ have the form (\ref{rhoA}).  We can now rewrite Eq.~(\ref{rhoAN}) as
\beq
\rho = \sum_h p(h) |A_j^{(N)}\rangle ... |A _i^{(1)}\rangle \langle A_i^{(1)}| ... \langle A_j^{(N)}|
= \sum_{f_\uparrow} p_N(f_\uparrow) \rho_N(f_\uparrow) .
\label{rhoAf}
\eeq
Here, the summation in the first step of Eq.~(\ref{rhoAf}) is over histories $h=\{ A_i^{(1)}, ..., A_j^{(N)} \}$ with $i,j = \uparrow, \downarrow$, the sum in the second step is over the values $f_\uparrow = 0, 1/N, 2/N, ..., 1$,  $p_N(f_\uparrow)$ is the binomial distribution (\ref{binomial}) expressed in terms of $f_\uparrow$, and $\rho_N(f_\uparrow)$ is the symmetrized and normalized density matrix for states with a given value of $f_\uparrow$.  For example, for $f_\uparrow = M/N$, where $M<N$ is an integer, 
\beq
\rho_N(f_\uparrow) = {\binom{N}{M}}^{-1} \left( |A_\downarrow^{(N)}\rangle ...  |A_\downarrow^{(M+1)}\rangle |A_\uparrow^{(M)}\rangle ... |A _\uparrow^{(1)}\rangle \langle A_\uparrow^{(1)}| ... \langle A_\uparrow^{(M)}|
\langle A_\downarrow^{(M+1)}| ... \langle A_\downarrow^{(N)}| + permutations \right) ,
\eeq
\beq
{\rm Tr} \rho_N(f_\uparrow) = 1.
\eeq

Now, in the limit $N\to\infty$, $p_N(f_\uparrow)$ becomes a
delta-function (\ref{delta}), so terms with $f_\uparrow \neq
|\alpha|^2$ drop out of the sum (\ref{rhoAf}), and only histories
satisfying the Born rule, $f_\uparrow =|\alpha|^2$, are represented in
the density matrix.  The final step of the argument, leading to the conclusion that quantum probabilities are given by the Born rule, is nontrivial and requires additional assumptions.  For example, one may have to assume \cite{Hartle} that a measurement of an observable ${\cal O}$ in a quantum state which is an eigenstate of ${\cal O}$ gives the corresponding eigenvalue with a 100\% probability.  It should also be noted that a careful definition of the limit $N\to\infty$ is a delicate issue, which is still being debated \cite{Farhi,Caves,Wesep,Wada}.  Here, I will not discuss these issues any further, since they are peripheral to the main subject of the present paper.



\section{Conclusions}

We discussed a possible extension of the watcher measure, introduced in \cite{watcher}, to quantum theory.  This measure identifies probabilities of different events in the multiverse with frequencies at which these events are encountered along the watcher's geodesic.  We have adopted a picture where the observable region of the watcher undergoes a stochastic evolution, due to its interaction with the environment.  The quantum state of this region is described by a reduced density matrix ${\tilde\rho}(t)$, where $t$ is a monotonic time variable along the watcher's worldline.  

A quantum extension of the watcher measure is most naturally obtained using the decoherent histories formulation of Quantum Mechanics.  In this formulation, histories are specified by chains of projection operators, and  the average number of events in a given time interval $T$ can be expressed in terms of a sum over histories, Eq.~(\ref{NA(T)}).  The relative probability of events is then given by the ratio of the corresponding occurrence numbers in the limit $T\to\infty$, Eq.~(\ref{pApB2}).  As in the classical version of the watcher measure, the resulting probabilities are independent of the choice of time variable $t$.  Note also that the ambiguity related to choosing between identical observers in an ensemble, pointed out in Refs.~\cite{Page,Srednicki}, does not arise in this measure, since there is no more than one observation that needs to be considered at any time. 

Any acceptable measure should be in agreement with the standard
predictions of Quantum Mechanics.  In particular, the probabilities of
possible outcomes of any quantum measurement should be given by the
expectation values of the corresponding projection operators (the Born
rule).  We have verified that this is indeed the case for the watcher
measure.  Furthermore, modulo the caveats associated with the
$N\to\infty$ limit, the Born rule can be derived from the watcher
measure in a manner similar to Refs.~\cite{Finkelstein,Hartle,Aguirre}. 

The watcher measure prescription gives a specific implementation of the idea that many worlds of the multiverse are the same as Everett's branching worlds \cite{Nomura,BoussoSusskind}.  Here, Everett's worlds are represented by the decoherent histories of the watcher, and the probabilities of different measurements in the multiverse are obtained from the watcher's density matrix.  As it now stands, this measure prescription is not fully quantum, since it relies on a semiclassical picture of spacetime and on classical concepts like the watcher's geodesic.  According to some approaches to quantum gravity, a full quantum description should be in terms of the wave function (or density matrix) of the causal patch of an observer (e.g., \cite{Nomura,census,BoussoSusskind}).  The background spacetime and geodesics representing possible trajectories of a watcher would then play no fundamental role, except perhaps in some appropriate limit.  Implementation of this approach, however, would require a better understanding of quantum gravity.

\subsection*{Acknowledgements}

I am grateful to Anthony Aguirre, Tom Banks, Ben Freivogel, Jaume Garriga, Alan Guth, Jim Hartle, Matt Kleban, Yasunori Nomura, Ken Olum and Don Page for very useful and stimulating discussions. 
This work was supported in part by the National Science Foundation (grant PHY-1213888) and by the Templeton Foundation.

\end{document}